\documentclass[10pt,aps,prd,twocolumn,preprintnumbers]{revtex4-1}

\usepackage{verbatim}
\usepackage{amsmath}
\usepackage{amsfonts}
\usepackage{amssymb}
\usepackage{bbm}
\usepackage{latexsym}
\usepackage{hhline}
\usepackage{color}
\usepackage{graphicx}
\usepackage{hyperref}

\newcommand{\aap}{Astron. Astroph.} 
\newcommand{\mnras}{MNRAS}
\newcommand{\apjl}{Astroph. J. Lett.}
\newcommand{\physrep}{Phys. Rep.}

\begin{document}

\title{A mechanism for fast radio bursts}

\author{G. E. \surname{Romero$^{1,2}$}}
%
\author{M. V. \surname{del Valle}$^1$}
\email[]{maria@iar-conicet.gov.ar}
\author{F. L. \surname{Vieyro}$^1$}
%
\affiliation{$^1$Instituto Argentino de Radioastronom\'{\i}a (IAR, CCT La Plata, CONICET), C.C.5, (1984) Villa Elisa, Buenos Aires, Argentina}
\affiliation{$^2$Facultad de Ciencias Astron\'omicas y Geof\'{\i}sicas, Universidad Nacional de La Plata, Paseo del Bosque s/n, 1900, La Plata, Argentina}

\begin{abstract}
Fast radio bursts are mysterious transient sources likely located at cosmological distances. The derived brightness temperatures exceed by many orders of magnitude the self-absorption limit of incoherent synchrotron radiation, implying the operation of a  coherent emission process.  We propose  a  radiation mechanism for fast radio bursts where the emission arises from collisionless Bremsstrahlung in strong plasma turbulence excited by  relativistic electron beams. We discuss possible astrophysical scenarios in which this process might operate. The emitting region is a turbulent plasma hit by a relativistic jet, where Langmuir plasma waves produce a concentration of intense electrostatic soliton-like regions (cavitons). The resulting radiation is coherent and, under some physical conditions, can be polarised and have a power-law distribution in energy. We obtain radio luminosities  in agreement with the inferred values for fast radio bursts. The timescale of the radio flare in some cases can be extremely fast, of the order of $10^{-3}$ s. The mechanism we present here  can explain the main features of fast radio bursts and is plausible in different astrophysical sources, such as  gamma-ray bursts and some Active Galactic Nuclei.
\end{abstract}

\pacs{} 
\keywords{}
%
%
\maketitle

\noindent

\section{Introduction}

Fast radio bursts (FRBs) are recently discovered transient sources of unknown origin (\cite{lorimer2007},~\cite{thornton2013},~\cite{spitler2014}). They were detected around $1.4$ GHz with a typical duration of $\lesssim  \delta t_{\rm{FRB}}\sim 10^{-3}$ s. Their location at high Galactic latitudes ($\left| b \right| > 40^{\circ}$) and  high dispersion measurements (DM$= 375-1103$ pc cm$^{-3}$) suggest propagation through the intergalactic medium and high redshifts. The observed radio fluences and the cosmological distances imply a total energy realise in radio waves of about $10^{40}$ erg and luminosities of $\sim 10^{43}$ erg s$^{-1}$. 

The extremely rapid variability points to relativistic beaming, so the linear size of the source would be $\delta x < c \Gamma^2 \delta t_{\rm FRB}$, where $\Gamma$ is the Lorentz factor of the source that is moving towards the observer \citep{piran1999}. The brightness temperatures associated with such compact and bright sources are extremely high: $T_{\rm b}> 10^{36} \Gamma^{-2}$ K (\cite{luan2014},~\cite{katz2014}). This is well above the Compton limit for incoherent synchrotron radiation. A coherent origin of the radiation, then, seems to be unquestionable. 

Additional constraints on the source can be obtained if we assume that the ultimate origin of the radiation is magnetic. A lower limit on the magnetic field that sets the particle flow in motion is $B^2 > 10^{19}\Gamma^{-3} (10^{-3} \rm{s} / \delta t_{\rm FRB})^{3}$ \citep{katz2014}. Even for large beaming, FRBs seem to be produced by compact objects of stellar origin such as neutron stars, magnetars or gamma-ray bursts (GRBs). In fact, a number of models have been proposed in relation to such objects: delayed collapses of supermassive neutron stars to black holes \citep{falcke2014}, magnetar flares \citep{popov2007}, mergers of binary white dwarfs \citep{kashiyama2013}, flaring stars \citep{loeb2014}, and short GRBs \citep{zhang2014}.  The radiation mechanism for the coherent emission is unknown.

In this article we propose that  FRBs are generated through coherent emission produced by a relativistic jet. Under some rather general conditions, beamed electrons interacting with self-excited strong turbulence produce collective radiation. The emission is generated by the interaction of the electrons with cavitons, which are the result of beam-excited Langmuir turbulence in a plasma traversed by the jet. This radiation mechanism, previously observed in laboratory experiments, has been studied in the context of intraday variability of Active Galactic Nuclei (AGN) jets \citep{benford1992}.

In the next section we present the basics of our model and show that it can explain the main features of FRBs. Then, in Sect. \ref{sources}, we discuss a possible astrophysical scenario where our proposed mechanism  might work. In Sect. \ref{discussion} we discuss our results, the problem of radio wave attenuation, { and the sensitivity of our model to the different physical parameters}. We close with a brief summary in Sect. \ref{conclusions}.

\section{Emission mechanism}\label{emi}

The interaction of a relativistic electron beam with a target  made out of plasma results, through  plasma instabilities,  in the generation of strong turbulence. This induced turbulence can be characterised as an ensemble of soliton-like wave packets, called \emph{cavitons} (\cite{zakharov1972},~\cite{wong1977},~\cite{kato1983}). These cavitons result from an equilibrium between the total pressure and the ponderomotive force, which causes a separation of electrons and ions. The cavity is then filled by a strong  electrostatic field $E_0$. This effect has been verified both in the lab (\cite{masuzaki1996},~\cite{robinson1997}) and through numerical simulations (\cite{sircombe2005},~\cite{henri2011}). 

Electrons passing through a caviton will radiate because they are accelerated by the electrostatic field, launching a broadband electromagnetic wave packet. The acceleration of the electrons of the  beam in the soliton field results in the superposition of the radiation of each electron. {The main contribution is produced by those cavitons with perpendicular  orientation which yields  perpendicular electron  acceleration.} If the beam is uniform, the out-coming  emission { is not coherent. However, when  some degree of inhomogeneity exists in the beam density, coherent radiation is produced.} The laboratory experiments clearly show that collective radiation processes occur when relativistic electrons are scattered by the cavitons; the electrons then produce a Bremsstrahlung-type of radiation of coherent nature  \citep{benford1992}. If a magnetic field is present, the cavitons can be aspherical, yielding {further circular polarised emission (even in the absence of a magnetic field some degree of polarization is expected due to random fluctuations of the size of the cavitons in the turbulent medium)}.

The condition for the collective radiatiation mechanism to operate is that the ratio of beam to plasma densities be no smaller than 0.01 \citep{benford1998}. This constraint arises in laboratory experiments that showed that bunching in the beam depends strongly on the ratio of beam to background plasma densities \citep{benford1992b}; a theoretical analysis for this effect is presented in \citet{benford1992}. For the case of a power-law density fluctuation spectrum (as expected in several turbulent regimes), the resulting radiation is also a power-law. The emerging spectrum of the emission is broadband, extending from the plasma frequency $\omega_e$ up to a cutoff around a few eV. In addition, the radiation is also relativistically beamed. For further details on the radiation process, readers are referred to \citet{weatherall1991} and \citet{benford1992}.

This emission mechanism has been discussed by \citet{benford1992b} in the context of intraday variability in quasars. Benford showed how the development of cavitons and coherent emission --extensively study in the laboratory-- can also take place in astrophysical sources regardless of the different scales involved. He argued that despite an astrophysical jet might have an electron energy distribution, i.e. a spread in $\gamma_{e}$, the jet velocity is always $\approx$ $c$ in the plasma frame, as in experiments, and hence the coherent emission is unavoidable if the approriate conditions in the plasma and beam are satisfied.

\subsection{Physical scenario}\label{scenario}

We propose that the interaction of a leptonic relativistic jet with a denser plasma cloud induces strong turbulence within the latter; the electrons then scatter with the cavitons producing radiation. The presence of density inhomogeneities in the jet causes the emission to be coherent. This latter condition is easily  fulfilled because there is always some level of density inhomogeneity in astrophysical fluids, given the high Reynolds number of the flows \citep[e.g.,][]{cho03}.

\begin{figure} 
\includegraphics[clip=true,width=.54\textwidth,angle=0]{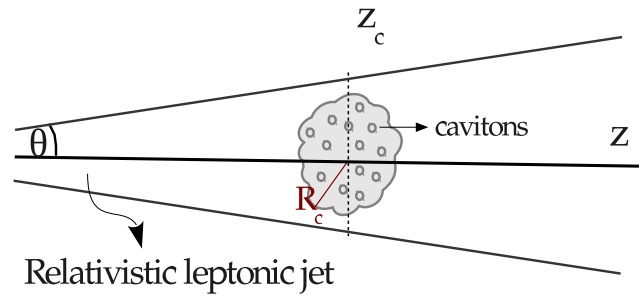}
\caption{Scheme of  the physical scenario considered in this work. Not to scale.}
\label{scheme}
\end{figure}

A sketch of the physical situation is presented in Figure~\ref{scheme}. In this system the required condition for collective emission is $n_{\rm j}/n_{\rm c} \ge 0.01$, with $n_{\rm j}$ and $n_{\rm c}$ the jet and cloud densities, respectively. The minimum radiation frequency { is set by the requirement that it must exceed the plasma frequency:} $\omega_e = 5.64 \times 10^4 n_{\rm{c}}^{1/2}$ Hz. For a density of $n_{\rm c} \sim 6 \times 10^{8}$ cm$^{-3}$ the plasma frequency is $\sim 1.4$ GHz, which corresponds to the observing frequency of FRBs.

For illustration purposes, we adopt the values listed in Table~\ref{Table 1} for the main parameters of our model (see Sect.~\ref{sources} for a discussion of the parameters).

\begin{table}
    \caption[]{Main parameters of the model.}\label{Table 1}
   	\label{table}
   	\centering
\begin{tabular}{ll}

\hline\hline %
Cloud parameters & Value\\ [0.01cm]
\hline

$n_{\rm{c}}$:   density [cm$^{-3}$] 								& $6 \times 10^{8}$\\
$T_{\rm{c}}$: 	temperature [K]				 							& $10^{5}$\\
$R_{\rm{c}}$: 	radius [cm]				 					 				& 5$\times 10^{13}$\\

\hline\hline 
Jet parameters & Value \\ [0.01cm]
\hline
$\Gamma$:       Lorentz factor                        & $500$\\
$n_{\rm{j}}$:   density [cm$^{-3}$] 								& $6 \times 10^{6}$\\

\hline  \\[0.005cm]
\end{tabular}	 	 																				
\end{table}

The size $D$ of the cavitons  induced into the cloud by the jet is $\sim 15\,\lambda_{\rm D}$ \citep{weatherall1991}, where ${\lambda_{\rm D}}$ is the Debye length of the plasma $\lambda_{\rm D}= 6.9\sqrt{T/n}$~cm. Then, 

\begin{equation}
D  \sim  1.3 \left(\frac{T_{\rm c}}{10^{5}\,{\rm K}}\right)^{1/2} \left(\frac{n_{\rm c}}{6 \times 10^{8}\,{\rm cm}^{-3}}\right)^{-1/2}\,\,{\rm cm},
\end{equation}

\noindent { where $T_{\rm c}$ is the temperature of the cloud of plasma (see Sect.~\ref{parameters} for further discussion, this temperature is ussually uncertain by an order of magnitude)}.

The impact of the jet on the denser plasma produces a shock that propagates in the direction of the jet motion and heats the cloud, which violently expands and is destroyed on timescales given by \citep[e.g.,][]{araudo2010}:

\begin{equation}\label{tclump}
t_{\rm c}\sim \frac{2 R_{\rm c} \sqrt{ \xi}}{c} \sim 3 \times 10^4 \left(\frac{R_{\rm c}}{5\times10^{13}\,{\rm cm}}\right)\,\,{\rm s},
\end{equation}

\noindent with $\xi = n_{\rm c}/ n_{\rm j}$, and $R_{\rm c}$ the radius of the cloud. This yields $t_{\rm c}$  $>>$ $\delta t_{\rm FRB}$. 

The pressure exerted by the jet will also accelerate the {cloud} to relativistic speed on a timescale of

 \begin{equation}\label{tg}
t_{\rm g}\sim \frac{ \xi R_{\rm c} }{c} \sim 10^5 \left(\frac{R_{\rm c}}{5\times10^{13}\,{\rm cm}}\right)\,\,{\rm s},
\end{equation}

\noindent which is also much longer than $\delta t_{\rm FRB}$.

The other fundamental timescale of the system is the time in which electrons radiate coherently; this scale is determined by the crossing time of the cloud, and dictates the duration of the event. In the laboratory frame, the crossing time is given by: 

\begin{equation}\label{tcool}
t_{\rm cross} = \frac{R_{\rm c}}{c\Gamma^2} \sim 6 \times 10^{-3} \left(\frac{R_{\rm c}}{5\times10^{13}\,{\rm cm}}\right)\left(\frac{500}{\Gamma}\right)^2 \,\,{\rm s};
\end{equation} 
  
Both dynamical timescales, {  $t_{\rm c}$ and $t_{\rm g}$}, are far longer than the duration of the FRB { given, in our model, by the crossing time $t_{\rm cross}$}. This means that the radiative phenomena, and not the dynamical disruption of the cloud,  are relevant to the evolution of the FRB. It is worth noticing that the crossing time and the clump radius are consistent with the upper limit imposed on the linear size of the source by the rapid variability.

Two shocks will be formed in the jet-cloud system that might re-accelerate particles. However, the timescale of the coherent losses of electrons is shorter than the timescale of acceleration even for relatively strong magnetic fields. For instance, if the magnetic field in the cloud has a  value of $10^{-2}$ G, we get $t_{\rm acc} \sim 10^{-2}$s. This means that in the presence of cavitons,  electron re-acceleration fails and they only emit coherently. Only protons can be efficiently accelerated in the cloud, but their emission through the $pp$ channel is too weak to be detectable.

\subsection{Radiated power}\label{results}

The radiated power per electron in the coherent region is given by \citep{weatherall1991}:

\begin{equation}\label{power}
	P = \frac{E_0 \sigma_{\rm{T}} c}{8 \pi}  \frac{4 n_{\rm{j}} \pi D^3}{3} \frac{27 \pi}{4} f 
	 \left[1 + \left( \frac{\Delta n_{\rm{j}}}{n_{\rm{j}}} \right)^2 0.24 \ln \left( \frac{2 \Gamma^2 c}{D \omega_e} \right)\right],	
\end{equation}

\noindent   where $\Delta n_{\rm{j}}/n_{\rm{j}}$ is the mean { squared} density fluctuation in the jet, { $E_0$ is the value of the electric field inside the cavitons}, and $f$ is the fraction of the cloud volume filled with cavitons. This fraction is a free parameter in our model; we adopt  $f \sim 0.1$, consistently with experiments that showed that $f$ can be as high as $0.5$ \citep{levron88}. We adopt density fluctuations of the order of 1\%, that is $\Delta n_{\rm{j}}^2/n_{\rm{j}}^2 \sim 10^{-4}$ \citep{weatherall1991}.

{
The value of $E_0$ can be estimated by the condition that allows the formation of the  cavitons, that is the electric energy must be greater than the thermal energy, i.e. ${E_0^2} / {8 \pi n_0 k_{\rm B}T_{\rm{c}}} << 1$ \citep{weatherall1991}. A  broad spectrum of density fluctuations will develop from the jet modulation through non-linear effects, of mean density $n_0$; we take  $n_0 = 10^3 n_{\rm{c}} $, as in \citet{weatherall1991}. We consider the energy ratio to be $0.1$, obtaining:

\begin{equation}
	E_0 \sim 4.6 \left(\frac{n_{\rm{c}}}{6 \times 10^{8}\,{\rm cm}^{-3}}\right)^{1/2} \left(\frac{T_{\rm{c}}}{10^5\,{\rm K}}\right)^{1/2}\,\,\,\,{\rm statV}~{\rm cm}^{-1}.
\end{equation}
}

The total power $P_{\rm t}$ is calculated as the power emitted per particle times the number of  relativistic electrons inside the cloud's volume, $N_e \sim 4/3 \pi \,n_{\rm j} R_{\rm c}^{3}$. This results in

\begin{equation}\label{power-total}
\begin{aligned} 
	P_{\rm t} & \sim  1.5 \times 10^{42} \,\,{\rm erg}\,{\rm s}^{-1} \\
	&\times \left(\frac{T_{\rm c}}{10^{5}\,{\rm K}}\right)^2 \left( \frac{f}{0.1} \right)
	\left(\frac{{n_{\rm j}}}{{6 \times 10^{6}\,{\rm cm}^{-3}}} \right) \left(\frac{{R_{\rm c}}}{{5\times10^{13}\,{\rm cm}}}\right)^{3}.
	\end{aligned}
\end{equation}

\noindent Here, we have considered $n_{\rm{c}} / n_{\rm{j}} \sim 100$, in accordance with the condition imposed by the radiative mechanism \citep{kato1983}.

The fluences of the detected bursts {at $\nu = 1.382$ GHz} are $0.6 - 8.0$~Jy~ms \citep{thornton2013}. Assuming an event of  average fluence $\sim$ 1~Jy~ms, with a duration of $1$~ms, at a distance of $2$~Gpc, the intrinsic luminosity is $\sim 5.8 \times 10^{42}$~erg~s$^{-1}$. For a source characterized by the parameters of Table \ref{table}, our model can account for the high fluences observed. In Sect. \ref{sources} we discuss possible astrophysical sources with these parameters. 

 As mentioned above, the coherent emission is broad band, extending from the plasma frequency $\omega_e$ up to $\nu_{\rm{max}} = 2 \gamma^2 c/D$. For $\gamma\sim \Gamma$, this yields:

\begin{equation}
\begin{aligned}
\nu_{\rm{max}} &\sim 1.1 \times 10^{7} \left(\frac{T_{\rm c}}{10^{5}\,{\rm K}}\right)^{-1/2}\\
&\times \left(\frac{n_{\rm c}}{10^{12}\,{\rm cm}^{-3}}\right)^{1/2} \left(\frac{\Gamma}{500}\right)^{2} \,\,{\rm GHz}.
\end{aligned}
\end{equation} 

\noindent Therefore $\nu_{\rm{max}} \sim 47.2$ eV, which is in the UV band. The absorption in the interstellar and intergalactic medium  suppresses the emission above the radio band. The radio flux, however, can be affected by plasma effects in the source (see Sect. \ref{discussion}).

\section{A possible astrophysical scenario: long gamma-ray bursts}\label{sources}

Recent bright GRBs detected with the \textsl{Fermi} satellite imply bulk Lorentz factors of $\Gamma \sim 1000$. These limits arise from the so-called ``compactness problem''. Such high Lorentz factors are necessary in order GeV photons to escape from the source (e.g., \cite{soderberg2003},~\cite{abdo2009},~\cite{ackermann2010},~\cite{ghisellini2010}). Even when applying more conservative models to describe the events,  values of $\gamma \sim 500$ are obtained \citep[e.g.,][]{chang2012}. GRBs involve, then, the fastest bulk motions known to occur in the Universe.

On the other hand, massive progenitor stars of long GRBs, such as  Wolf-Rayet stars, have strong  winds with a clumpy structure (e.g., \cite{hillier2003},~\cite{owocki2006}). Once the star implodes, one or more clumps can be reached by the jet, because of the relative high filling factor \citep{owocki2009}. The interaction of clumps and/or clouds with jets, along with strong turbulence  generation, produces different phenomena such as particle acceleration, non-thermal emission, etc. (see \cite{araudo2009},~\cite{araudo2010}). We propose here that the interaction of a long GRB jet with the clumped, residual wind of its progenitor can lead to the coherent emission previously discussed. 
 
The density of a clump in the wind of a Wolf-Rayet star can reach very high values, similar to those expected in the atmosphere of massive stars ($n_{\rm c} \sim 10^{11-12}$ cm$^{-3}$, \citep{Crowther2008},~\cite{araudo2009}). As these clumps move away from the star, they expand and their density decrease to values similar to the ones adopted in our model. On the other hand, the jet density is determined by the luminosity of the GRB, and it also decreases with the distance to the central source. For typical GRB luminosities of $\sim 10^{50}$ erg s$^{-1}$, and interaction distances of $10^{15-16}$ cm (this is located near the region where the afterglow emission is produced), the density can easily reach values of $n_{\rm j} \sim 0.01 n_{\rm c}$, which are necessary for the proposed mechanism to operate. Since the interaction between the clump and the jet should occur far enough to the central engine, the radio coherent emission is not screened neither by the dense stellar envelope nor the GRB prompt emission.

Finally, we note that the fraction $q$ of the jet kinetic energy of a GRB transferred into the collective emission is $q \sim 10^{-7}$; this is a rather slim value of $q$, considering the theoretical upper limits obtained in \citet{benford1992}, that give $q \sim 10^{-3}-0.5$. The energetic requirements for this implementation of our mechanism are, then, not very demanding.

\section{Discussion}\label{discussion} 

Coherent radiation processes have been claimed to suffer severe attenuation by various absorption mechanisms in the context of AGNs (e.g., \cite{levinson95},~\cite{benford1998}). The brightness temperature  can be  saturated by induced Compton scattering and/or Raman scattering (\citep{coppi93},~\cite{levinson95}). However, \citet{benford1998} argued that confrontation of induced Compton absorption with plasma experiment suggests that there is no observed saturation effect for high $T_{\rm B}$. Regarding Raman scattering, the theoretical approach  in \citet{levinson95} applies only to weakly turbulent environments. An order of magnitude estimate, however, can be obtained.  

Strong Raman scattering dominates other decay processes of Langmuir waves if  
\citep{levinson95}:

\begin{equation}\label{uno}
\left(  \frac{n_{6} T_{\rm B12}}{\nu_{9}^{2}}\right) > 2\times 10^5 \,,
\end{equation} 

\noindent where $n_6$ is the plasma density in units of $10^6$ cm$^{-3}$, $\nu_9$ is the frequency of the radio emission is GHz, and $T_{B12}$ is the brightness temperature in units of $10^{12}$ K. 
  
The brightness temperature $T_{\rm B}$ can be calculated as follows \citep[e.g.,][]{luan2014}:
 
\begin{eqnarray}
T_{\rm B} &\sim &\frac{S_{\nu}\,d^{2}}{k_{\rm B}\,{\gamma}^2 {\nu}^2{\Delta}t^2}\\ 
&\sim&  \frac{3.5 \times 10^{34}\,{\rm K}}{\Gamma^2}\left(\frac{S_{\nu}}{\rm Jy}\right)\left(\frac{d}{\rm Gpc}\right)^{2}\left(\frac{\nu}{\rm GHz}\right)^{-2}\left(\frac{\Delta t}{\rm ms}\right)^{-2}. \nonumber
\end{eqnarray}
 
For a typical burst, $T_{\rm B} \sim 1.4 \times 10^{29}$~K. With these values condition (\ref{uno})  holds. However, experiments can produce effective brightness temperatures in excess of $10^{30}$ K using laser-like devices to stimulate a hot plasma \citep{robinson1997, benford1998}. This discrepancy between theory and experiment might arise in the fact that the above calculation is based on the weak-turbulence limit, which is not adequate to describe the situation under consideration here. We conclude that there is no reason, in the absence of a comprehensive theory of strong Langmuir turbulence, to rule out our model given the experimental results.    

In our model we consider that a cloud of plasma interacting with the relativistic jet produces the required density rarefaction for the generation of cavitons; however regions of very high density in the jet can be formed by other processes, like internal shocks, instabilities, etc. In these cases the coherent emission resulting from electron-caviton scattering might also be obtained. 

A different scenario might be associated with the minijets that are  produced by  dissipation of magnetic energy in a larger jet \citep{giannios2009}. For instance, in the jet of a misaligned AGN, a minijet pointing towards the observer can impact with a cloud from the broad line region (BLR). These clouds have high densities, typical sizes of $R_{\rm c} \sim 10^{13}$ cm \citep{risaliti2010} and temperatures $T_{\rm c} \sim 2\times10^{4}$K \citep{araudo2010}. The jets of AGNs have Lorentz factors $\Gamma_{\rm j} \sim 10$, however minijets might be much faster (Lorentz factors $\Gamma_{\rm mj} >> \Gamma_{\rm j}$, with $\Gamma_{\rm mj} \sim$ 200, \citep{giannios2009}). With these parameters the values for $t_{\rm cross}$ and the total power do not differ significantly from those obtained in the case discussed in Sect.\ref{emi}. 

Beam decollimation might occur in astrophysical jets. Decollimation into an angle $\Phi$ will not affect the collective emission until $\Phi > 1/\Gamma$, but then emission will drop greatly if $\Phi \sim \pi/2$ \cite{benford1992b}. Such high angles are not expected in the  astrophysical scenario presented here.

Multiple FRBs from one source depend on the filling factor of clouds. For a value of 0.1, the probability of having multiple independent interactions is $\sim$ $10^{-2}$. { Even if there is a multiple interaction, it would appear as a bit longer FRB ($t \sim 10^{-2}$~s) with some structure, as recently observed  \cite{champion15}}.

In a recent FRB detected in real time, Petroff and co-workers \cite{petroff2015} have measured circular polarisation of $21\pm 7$\%, and established an upper limit on linear polarization of $<10$ \% (with $1\sigma$ of significance). There are at least two ways of producing circular polarization: 1) in a jet composed by electron/positron pairs, circular polarisation might be due to Faraday conversion (\cite{jones1988},~\cite{wardle1998}); 2) in case that some cyclotron modes scatter with the jet, higher levels of circular polarisation are expected. Both effects can contribute to the high degree of circular polarisation reported by \citet{petroff2015}. The radiation mechanism we are proposing might produce linear polarisation as well \citep{benford1998}. The low levels of linear polarisation detected may be due to Faraday rotation as proposed in \citet{petroff2015}. 

{ Effects of external plasma, such as propagation effects (see e.g., \cite{katz2015}), can be also invoked to explain the existence of some spectral features claimed to be present  (e.g., bright bands of a width $\sim 100$ MHz, \cite{thornton2013}).} 

{ The first evidence of a two-component FRB disfavors models that resort to a single high energy event involving compact objects (e.g., \cite{champion15}). Our model, on the contrary, can easily explain two components or a structured time profile by multiple interactions of different clumps with the jet. It can also account for two FRBs being produced by the same repeating source, as suggested by \cite{maoz2015}. Furthermore, recent investigations conclude that FRBs arise in dense star-forming regions (e.g., \cite{kulkarni15},~\cite{masui15}), precisely where massive stars with clumpy winds are expected.}

{Another aspect to be briefly discussed is the potential correlation between long GRBs and FRBs. No such a correlation has been observed so far, but with only around 10 events detected this is not surprising at all. Notice that the interaction might occur at some distance from the region where the gamma-rays are produce, hence the GRB and the FRB are not exactly simultaneous. In addition, the deceleration of the jet makes the beaming angle of the radio emission larger than that of the gamma-ray flux, making the GRBs more difficult to be detected in comparison to the FRBs. Finally, not all GRBs are expected to gather the adequate environment or jet conditions for producing the FRB phenomenon. }

\subsection{Sensitivity to model parameters}\label{parameters} 

{ The luminosity and event duration in our model depend on the characteristics of the target plasma (clump)--these are properties such as clump size, density, and temperature-- and the jet parameters --density and Lorentz factor--. Here we discuss the sensitivity of our results to  these parameters. }

{
The detection frequency of FRBs imposes an upper limit to the density of the cloud in our model (as discussed in Sect.~\ref{scenario}); hence denser clumps might emit at higher frequencies, beyond the radio band, and are discarded. The density value we adopted is reasonable for a clump of the wind of a massive star (such as a Wolf-Rayet).}

{In the application of our model we have adopted a density ratio $n_{\rm j}/n_{\rm c}=0.01$;  the required condition for collective emission is $n_{\rm j} /n_{\rm c} \geq 0.01$,  since $0.01$ is a lower limit to the density ratio,  bigger values can be adopted. If we consider, for example, $n_{\rm j}/n_{\rm c}=0.1$, we obtain detectable fluxes for smaller clumps ($R_{\rm c}=5\times 10^{11}-10^{12}$ cm), and jets with lower Lorentz factors ($\Gamma \sim 80 - 100$). Clumps smaller than $10^{11}$~cm might produce undetectable fluxes.
}

{Clumpy structures in the wind of massive stars develop close to the surface of the star, with typical sizes of $\sim 10^{11}$~cm. We have adopted a larger clump's size of $5\times 10^{13}$~cm. This value is justified by the fact that the clump-jet interaction in our model takes place far from the central source, and at such distances the clump is expected to naturally expand. In addition, in the later stages before the final collapse of evolved  stars, massive ejections occur from the outer atmosphere, as observed in the case of Eta Carina. }

{We have adopted a temperature of the clump of $10^5$~K, which corresponds to the temperature of massive star winds \citep{krticka2001}. This is a very conservative value, since stellar winds are known to emit even soft X-rays. This implies that the plasma can reach temperatures of $T=10^{6-7}$~K (\cite{oskinova2011}, \cite{ignace2012}). Incrementing the temperature one order of magnitude increases approximately one order of magnitude the resulting luminosity; in this way, by adopting a hotter clump other parameters can be relaxed, such as the clump's size and/or the jet's Lorentz factor.}

{We conclude, then, that there is a wide range of parameters over which our model is able to reproduce the main features observed so far in FRBs. 

}

\section{Conclusions}\label{conclusions}

We propose a model where FRBs are the result of the interaction of relativistic jets with plasma condensations. In this context, the jet induces strong turbulence in the cloud, exciting Langmuir waves. In the strong turbulence regime cavitons are formed, filling a substantial part of the volume of the cloud. In the presence of an inhomogeneous jet,  these cavitons coherently scatter electrons with Lorentz factors of $\sim 100$ producing radiation up to a frequency of $\sim 10^4 c D^{-1}$, with $D$ the size of the cavitons. The result is a radio flare with an effective brightness temperature exceeding $10^{32}$~K, and a duration of $\sim 10^{-3}$ s in the observer reference system. If the jet density fluctuations have a power-law distribution, the resulting radiation can have a power-law spectra as the synchrotron radiation, but it is not restricted by the high brightness temperatures derived assuming incoherent emission. We suggest that these events can account for at least a fraction of the FRBs. Several possible scenarios can accommodate this mechanism. Among them, we briefly discussed jets of long GRBs colliding with inhomogeneities of the wind of the progenitor star or the warm component of the ISM. Other possible scenarios for the proposed mechanism will be discussed elsewhere. 

Polarisation measurements \citep[e.g.,][]{petroff2015} can shed light on the magnetic fields present in the clouds and the deformation of the cavitons.  

\vskip.5cm 
\acknowledgments
This work was supported by the Argentine Agencies CONICET  and ANPCyT (PICT 2012-00878), as well as by grant AYA2013-47447-C3-1-P (Spain). It should have been also supported by an unpaid CONICET PIP grant (2014) to G.E.R.


\end{document}